\documentclass[aps,prb,superscriptaddress,twocolumn,showpacs,amsmath,amssymb]{revtex4-1}
\usepackage{stmaryrd}
\usepackage{times}
\usepackage{graphicx}
\usepackage{dcolumn}
\usepackage{bm}

\begin{document}

\title{Observation of pure inverse spin Hall effect in ferromagnetic metals by FM/AFM exchange bias structures}

\author{H. Wu}
\affiliation{ Beijing National Laboratory for Condensed Matter Physics, Institute of Physics, Chinese Academy of Sciences, Beijing 100190, China}

\author{C. H. Wan}
\email[Email: ]{wancaihua@iphy.ac.cn}
\affiliation{ Beijing National Laboratory for Condensed Matter Physics, Institute of Physics, Chinese Academy of Sciences, Beijing 100190, China}

\author{Z. H. Yuan}
\affiliation{ Beijing National Laboratory for Condensed Matter Physics, Institute of Physics, Chinese Academy of Sciences, Beijing 100190, China}

\author{X. Zhang}
\affiliation{ Beijing National Laboratory for Condensed Matter Physics, Institute of Physics, Chinese Academy of Sciences, Beijing 100190, China}

\author{J. Jiang}
\affiliation{ Beijing National Laboratory for Condensed Matter Physics, Institute of Physics, Chinese Academy of Sciences, Beijing 100190, China}

\author{Q. T. Zhang}
\affiliation{ Beijing National Laboratory for Condensed Matter Physics, Institute of Physics, Chinese Academy of Sciences, Beijing 100190, China}

\author{Z. C. Wen}
\affiliation{ Beijing National Laboratory for Condensed Matter Physics, Institute of Physics, Chinese Academy of Sciences, Beijing 100190, China}

\author{X. F. Han}
\email[Email: ]{xfhan@iphy.ac.cn}
\affiliation{ Beijing National Laboratory for Condensed Matter Physics, Institute of Physics, Chinese Academy of Sciences, Beijing 100190, China}

\begin{abstract}

We report that the spin current generated by spin Seebeck effect (SSE) in yttrium iron garnet (YIG) can be detected by a ferromagnetic metal (NiFe). By using the FM/AFM exchange bias structure (NiFe/IrMn), inverse spin Hall effect (ISHE) and planar Nernst effect (PNE) of NiFe can be unambiguously separated, allowing us to observe a pure ISHE signal. After eliminating the in plane temperature gradient in NiFe, we can even observe a pure ISHE signal without PNE from NiFe itself. It is worth noting that a large spin Hall angle (0.098) of NiFe is obtained, which is comparable with Pt. This work provides a kind of FM/AFM exchange bias structures to detect the spin current by charge signals, and highlights ISHE in ferromagnetic metals can be used in spintronic research and applications.

\end{abstract}


\maketitle


How to generate, manipulate, and detect spin currents ({$\textbf{\emph{J}}_{\mbox{\tiny S}}$}) is a fundamental issue in spintronic research \cite{1Wolf2001,2uti2004}. Spin injection from a ferromagnetic metal \cite{3Johnson1985,4jedema2001electrical}, spin pumping \cite{5saitoh2006conversion,6mosendz2010quantifying}, spin Hall effect (SHE) \cite{7kato2004observation,8valenzuela2006direct}, and spin Seebeck effect (SSE) \cite{9Uchida2008,10bauer2012spin,11uchida2010spin,12uchida2010observation,13uchida2010longitudinal,14li2014generation,Adachi2013} provide several ways to generate a spin current. Especially SSE in ferromagnetic insulators (FI) \cite{11uchida2010spin,12uchida2010observation,13uchida2010longitudinal,14li2014generation} has attracted much attention for a pure spin current can be generated without any charge flow. Inverse spin Hall effect (ISHE) \cite{5saitoh2006conversion,15kimura2007room} in heavy metals with strong spin-orbit coupling (SOC) such as Pt is often used to detect the spin current by charge signals: $\textbf{\emph{E}}_{\mbox{\tiny ISHE}}=(\theta_{\mbox{\tiny SH}}\rho)\textbf{\emph{J}}_{\mbox{\tiny S}} \times \bm\sigma$, where $\textbf{\emph{E}}_{\mbox{\tiny ISHE}}$ is the ISHE electric field, $\theta_{\mbox{\tiny SH}}$ is the spin Hall angle, $\rho$ is the resistivity and  $\bm\sigma$ is the unit vector of spin.

As the inverse effect of anomalous Hall effect (AHE), ISHE in ferromagnetic metals provides a possibility to detect the spin current as well. Recently, several works focus on using ferromagnetic metals instead of metals with strong SOC to detect the spin current generated by SSE in FI \cite{16miao2013inverse,17Tian2015Separation,18wu2014unambiguous}. However, additional anomalous Nernst effect (ANE) and planar Nernst effect (PNE) in the ferromagnetic metal itself is often mixed with the ISHE signal in longitudinal and transversal spin Seebeck measurement respectively. Therefore, in transversal spin Seebeck measurement, unambiguous separation of PNE and ISHE signals will be an important progress, not only for exploring the physical mechanism of ISHE in ferromagnetic metals, but also for future applications in detecting spin currents.

Exchange bias phenomenon in the ferromagnetic (FM)/antiferromagnetic (AFM) interface \cite{19koon1997calculations,20berkowitz1999exchange} can provide a shift field ($H_{\mbox{\tiny EB}}$) of the magnetization hysteresis loop, when cooling down to Neel temperature ($T_{\mbox{\tiny N}}$) with a static magnetic field, which has been used in spin valve structures for several years. This phenomenon is associated with the interfacial exchange anisotropy between FM and AFM, and FM tends to align parallel with uncompensated spins of AFM at the interface. Therefore, FM has a unidirectional anisotropy.

In this work, NiFe/IrMn exchange bias structure has been employed to detect the spin current in NiFe originating from SSE in YIG, Cu was inserted between NiFe and YIG to decrease the exchange coupling and to eliminate the possible magnetic proximity effect \cite{21huang2012transport,22lu2013pt}. The temperature gradient  $\nabla T$ is mainly in plane and along the exchange bias field axis. However, PNE from NiFe itself will be involved in ISHE voltages \cite{23pu2006anisotropic,24avery2012observation}. This structure can separate the magnetization reversal process of YIG and NiFe. As a result, ISHE and PNE which related to the magnetization state of YIG and NiFe respectively could be separated as well.

The detail multilayer film structure is GGG/YIG/Cu(t nm)/NiFe(5 nm)/IrMn(12 nm)/Ta(5 nm). Firstly, a 3.5 $\mu$m YIG film was grown on a 300 $\mu$m GGG(111) substrate using liquid phase epitaxial method. Then upper films were deposited using an ultrahigh vacuum magnetron sputtering system (ULVAC) at a pressure of 0.16 Pa and a power of 120 W. In order to provide a clear interface between YIG and Cu, the YIG surface was cleaned for 60 s by Ar plasma in the vacuum chamber before deposition. A 100 Oe magnetic field was applied during deposition, which could induce an easy magnetization axis and an exchange bias of NiFe. Films were patterned by photolithography combined with Ar ion etching. Both of the electrodes A and C are of 10 $\mu$m $\times$ 100 $\mu$m in size, and the size of electrode B is 50 $\mu$m $\times$ 100 $\mu$m ($L=100 \mu$m). The spacing between A (B) and B (C) is 10 $\mu$m.

Fig. 1(a) shows the schematic illustration of the measurement method. Electrode A and C were used to heat the YIG film by electric currents  $I_{\mbox{\tiny H}}$ (Keithley 2440), which induced a transverse temperature gradient $\nabla T$ mainly along \emph{y} axis, and the heating power $P\propto I^{2}_{\mbox{\tiny H}}\propto \nabla T$ . Because of SSE in YIG, $\nabla T$  produces a spin accumulation at the interface between YIG and electrode B, and then the spin current is injected to electrode B. By measuring the voltage along \emph{x} axis in electrode B (Keithley 2182A), the spin current can be detected by means of ISHE, as shown in Fig. 1(b). The physical property measurement system (Quantum Design PPMS) was used to apply the magnetic field and control the temperature. All measurements were performed at room temperature.

\begin{figure*}
\includegraphics[width=130mm]{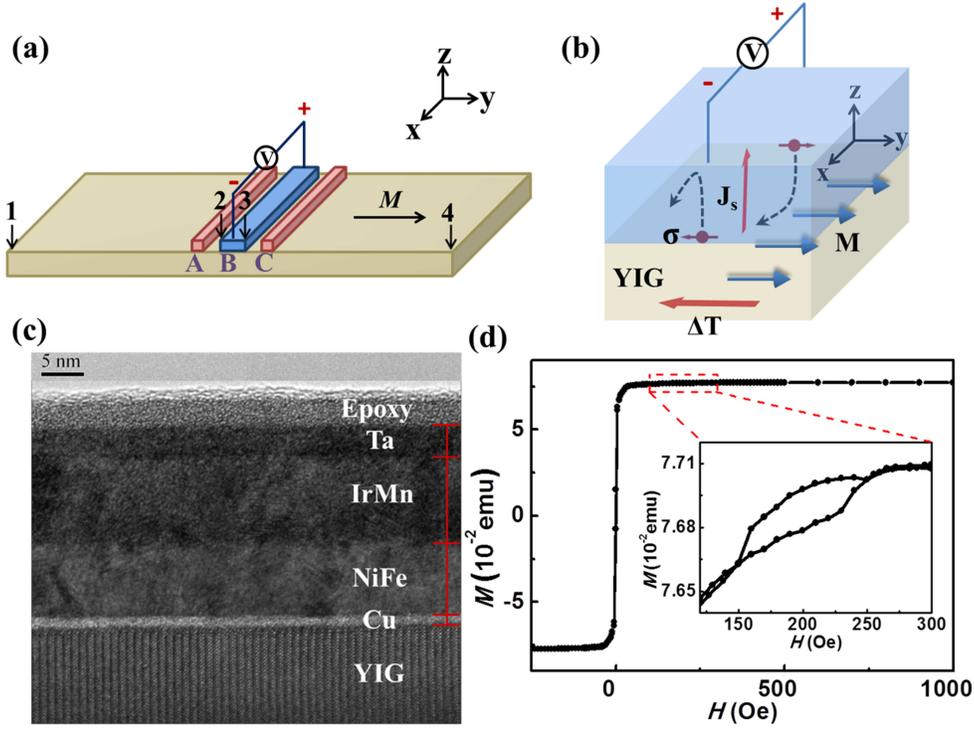}
\caption{\label{fig1} (a) A schematic of patterned device structures, A and C electrodes are for heating currents $I_{\mbox{\tiny H}}$ and B electrode is for ISHE voltages $V_{\mbox{\tiny ISHE}}$  measurement. (b) A schematic illustration of ISHE in electrode B induced by SSE in YIG. The temperature gradient $\nabla T$ is mainly along \emph{y} axis and the spin current in B is along \emph{z} axis, therefore the ISHE voltage is measured along \emph{x} axis. (c) Cross section HRTEM results of YIG/Cu(3 nm)/NiFe(5 nm)/IrMn(12 nm)/Ta(5 nm) sample for detecting the spin current. (d) \emph{M}-\emph{H} loops of YIG/Cu(5 nm)/NiFe(5 nm)/IrMn(12 nm)/Ta(5 nm) sample, the magnetic field is along \emph{y} axis. }
\end{figure*}

The cross-section high resolution transmission electron microscopy (HRTEM) of GGG/YIG/Cu(3 nm)/NiFe(5 nm)/IrMn(12 nm)/Ta(5 nm) sample was observed by Tecnai G2 F20 S-TWIN (200 kV). HRTEM results are shown in Fig. 1(c). The high quality YIG single crystal structure is formed on the GGG(111) substrate, and the epitaxial direction of YIG film is also along (111) direction. Four metal layers deposited by magnetron sputtering are continuous and flat, and each interface especially the interface between YIG and Cu is very clear and sharp. The spin current is injected from YIG to above films, so the clear YIG/Cu interface is very important.

The magnetic hysteresis loop of GGG/YIG/Cu(5 nm)/NiFe(5 nm)/IrMn(12 nm)/Ta(5 nm) sample was measured by a vibrating sample magnetometer (VSM, MicroSense EZ-9) with magnetic field applied along \emph{y} axis (also the axis of the exchange bias field), as shown in Fig. 1(d). YIG is a very soft magnetic material and the saturation field ($H_{\mbox{\tiny S}}$) of YIG is less than 10 Oe. The inserted figure shows the minor \emph{M}-\emph{H} loop from NiFe, and $H_{\mbox{\tiny EB}}$ (200 Oe) is enough to distinguish the magnetization reversals of NiFe and YIG. Besides, the magnetic moment from YIG is very large due to its larger thickness.

As reported in previous works \cite{9Uchida2008,10bauer2012spin,11uchida2010spin,12uchida2010observation,13uchida2010longitudinal,14li2014generation}, firstly we used a 10 nm thick ($d_{\mbox{\tiny Pt}}$) Pt film to detect {$\textbf{\emph{J}}_{\mbox{\tiny S}}$} induced by SSE in YIG. A 300 nV ISHE voltage is observed as $I_{\mbox{\tiny H}}=10$ mA in electrode C is applied with field along \emph{y} axis [Fig. 2(a)]. ISHE voltages were not observed when field was applied along \emph{x} and \emph{z} axis respectively, which confirms the SSE scenario. When a 3 nm metal Cu layer is inserted between Pt and YIG to eliminate the magnetic proximity effect between YIG and Pt, still a spin current can pass without remarkable dissipation, as proven by the ISHE voltage observed in this case. However, once a 3 nm insulator MgO layer is inserted to block {$\textbf{\emph{J}}_{\mbox{\tiny S}}$} from YIG, the ISHE voltage completely disappears. These results confirm that the voltage is induced by {$\textbf{\emph{J}}_{\mbox{\tiny S}}$}  injected from YIG. This voltage does not come from Pt or YIG alone, which could be proven by the absence of the voltage in YIG/Cu and Si-SiO$_2$/Pt reference samples.

\begin{figure*}
\includegraphics[width=130mm]{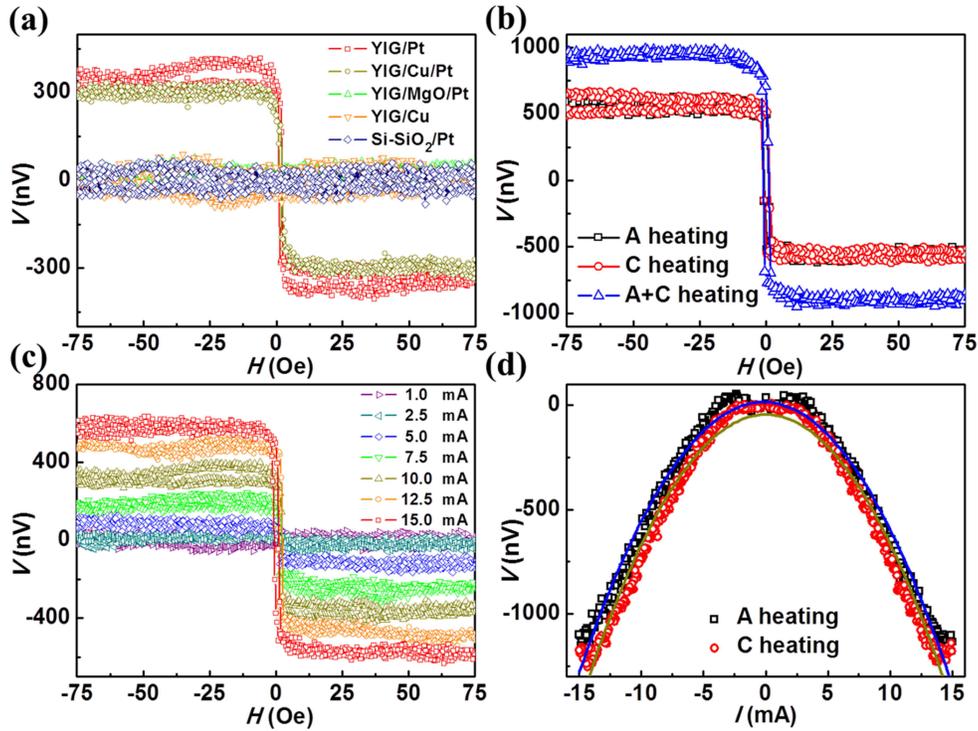}
\caption{\label{fig2} (a) \emph{H} dependence of $V_{\mbox{\tiny ISHE}}$ in YIG/Pt(10 nm), YIG/Cu(3 nm)/Pt(10 nm), YIG/MgO(3 nm)/Pt(10 nm), YIG/Cu(10 nm), Si-SiO$_2$/Pt(10 nm) samples. (b) \emph{H} dependence of $V_{\mbox{\tiny ISHE}}$ for heating A or C respectively, and simultaneously heating A and C in YIG/Pt(10 nm) sample. (c) \emph{H} dependence of $V_{\mbox{\tiny ISHE}}$  for different $I_{\mbox{\tiny H}}$ in electrode C in YIG/Pt(10 nm) sample. (d) $I_{\mbox{\tiny H}}$ dependence of $V_{\mbox{\tiny ISHE}}$ and fitting curves for heating A or C in YIG/Pt(10 nm) sample.}
\end{figure*}

When we changed the heating electrode from C to A: $T_{\mbox{\tiny B,A}}$, $T_{\mbox{\tiny 1,A}}$, $T_{\mbox{\tiny 4,A}}$ and $T_{\mbox{\tiny B,C}}$, $T_{\mbox{\tiny 1,C}}$, $T_{\mbox{\tiny 4,C}}$ represent the temperature of point B, 1, 4 when heating A and C respectively; $T_{\mbox{\tiny B,A+C}}$ , $T_{\mbox{\tiny 1,A+C}}$  and  $T_{\mbox{\tiny 4,A+C}}$ represent the temperature of point B, 1 and 4 when heating A and C simultaneously. $T_{\mbox{\tiny 1,C}}=T_{\mbox{\tiny 4,A}}$, $T_{\mbox{\tiny B,A}}=T_{\mbox{\tiny B,C}}$, $T_{\mbox{\tiny 1,A+C}}=T_{\mbox{\tiny 4,A+C}}$ due to the geometrical symmetry. So the ISHE voltage: $V_{\mbox{\tiny ISHE,A}}=S_{\mbox{\tiny 1}}(T_{\mbox{\tiny B,A}}-T_{\mbox{\tiny 4,A}})=V_{\mbox{\tiny ISHE,C}}=S_{\mbox{\tiny 1}}(T_{\mbox{\tiny B,C}}-T_{\mbox{\tiny 1,C}})$ , where $S_{\mbox{\tiny 1}}=\frac{1}{2}\theta_{\mbox{\tiny Pt}}\eta_{\mbox{\tiny YIG-Pt}}(L_{\mbox{\tiny Pt}}/d_{\mbox{\tiny Pt}})S_{\mbox{\tiny S}}$, $\theta_{\mbox{\tiny Pt}}$  is the spin Hall angle of Pt,  $\eta_{\mbox{\tiny YIG-Pt}}$ is the spin injection efficiency, $L_{\mbox{\tiny Pt}}/d_{\mbox{\tiny Pt}}$  is the aspect ratio and  $S_{\mbox{\tiny S}}$ is the spin Seebeck coefficient \cite{9Uchida2008}. The ISHE voltage is almost the same when changing the heating electrode from C to A, as shown in Fig. 2(b). When heating A and C at the same time, the ISHE voltage is enhanced due to higher temperature gradient: $V_{\mbox{\tiny ISHE,A+C}}=S_{\mbox{\tiny 1}}(T_{\mbox{\tiny B,A+C}}-T_{\mbox{\tiny 4,A+C}})=S_{\mbox{\tiny 1}}(T_{\mbox{\tiny B,A+C}}-T_{\mbox{\tiny 1,A+C}})$  [Fig. 2(b)].

We also measured the  $V_{\mbox{\tiny ISHE}}-I_{\mbox{\tiny H}}$ curves with fields along \emph{y} axis larger than $H_{\mbox{\tiny S}}$ of YIG ($\pm$ 20 Oe) and then obtained the difference between them, namely spin dependent ISHE voltages: $V_{\mbox{\tiny ISHE}}=V(+M_s)-V(-M_s)$. Fig. 2(c) and Fig. 2(d) show the relationship between ISHE voltages and heating currents: $V_{\mbox{\tiny ISHE}}\propto I^2_{\mbox{\tiny H}}\propto \nabla T$ , which confirms that the ISHE signal is thermal related. And the $V_{\mbox{\tiny ISHE}}-I_{\mbox{\tiny H}}$ curves nearly coincide after changing the heating electrode from C to A.

Furthermore, we changed the spin current detector Pt with the exchange bias structure: Cu(5 nm)/NiFe(5 nm)/IrMn(12 nm)/Ta(5 nm), and heated the electrode C with $I_{\mbox{\tiny H}}=$15 mA . The heating current generates $\nabla T$ not only in YIG, but also in electrode B, which induces a PNE voltage in NiFe. By using the exchange bias structure, magnetization reversals of NiFe and YIG are separated, as can be seen in Fig. 1(d). As a result, ISHE (related to magnetization of YIG) and PNE (related to magnetization of NiFe) are separated as well. As shown in Fig. 3(a), a 500 nV PNE voltage is observed and the center field of the PNE curve locates at 120 Oe. This shift field is smaller than the $H_{\mbox{\tiny EB}}$ from \emph{M}-\emph{H} curves for two reasons: one is that the film is patterned, and another is that the temperature of the electrode B increases when heating C.

\begin{figure*}
\includegraphics[width=130mm]{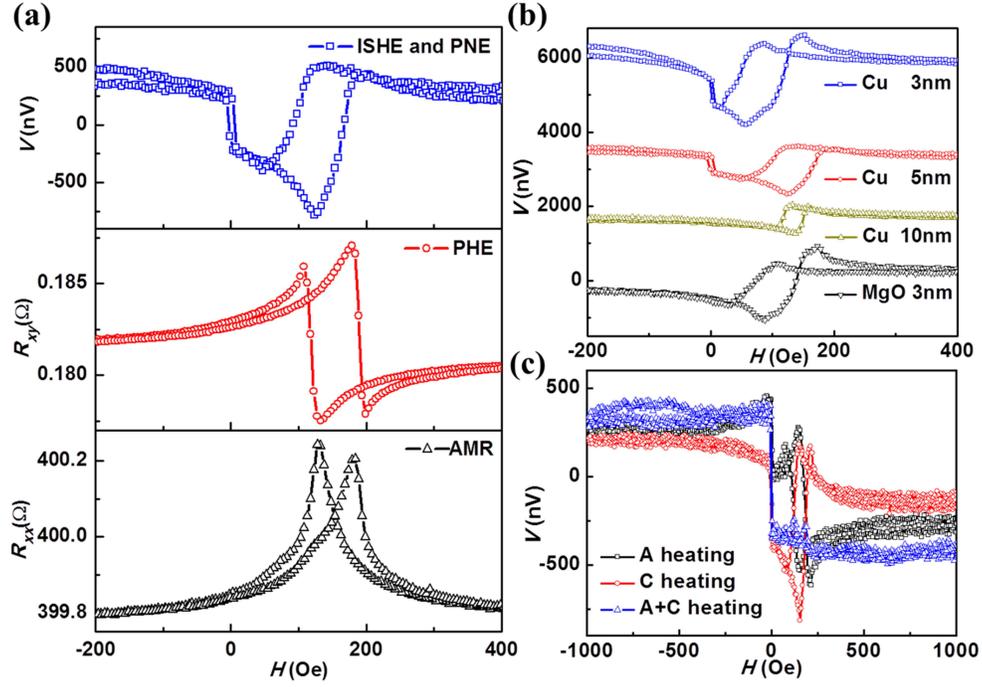}
\caption{\label{fig3}(a) \emph{H} dependence of ISHE, PNE, AMR and PHE signals in YIG/Cu(5 nm)/NiFe(5 nm)/IrMn(12 nm)/Ta(5 nm) sample. (b) \emph{H} dependence of  $V_{\mbox{\tiny ISHE}}$ and $V_{\mbox{\tiny PNE}}$ in YIG/x/NiFe(5 nm)/IrMn(12 nm)/Ta(5 nm) samples with different inserted layers, the inserted layer x = Cu 3 nm, Cu 5 nm, Cu 10 nm, MgO 3 nm. (c) \emph{H} dependence of $V_{\mbox{\tiny ISHE}}$ and $V_{\mbox{\tiny PNE}}$ for heating A or C respectively, and simultaneously heating A and C in YIG/Cu(5 nm)/NiFe(5 nm)/IrMn(12 nm)/Ta(5 nm) sample.}
\end{figure*}

It is especially attractive that a 250 nV $V_{\mbox{\tiny ISHE}}$  is observed near zero magnetic field and the voltage saturates at a field less than 10 Oe, which is similar to the signal in YIG/Pt sample. And the sign of the ISHE voltage in NiFe is the same with that in Pt. Transport properties only depend on the magnetization of NiFe, because YIG is an insulator. Anisotropic magnetoresistance (AMR) and planar Hall effect (PHE) reflect the magnetization state of NiFe and share the similar origin with PNE, which only have a signal near 150 Oe, and do not have an obvious signal near 0 Oe. Especially PHE almost have the same curve with PNE, the only difference is that one is from the electric current, and the other is from the thermal current. These prove that the signal near 0 Oe is not from PNE in NiFe, but from ISHE in NiFe induced by SSE in YIG, which can also be confirmed by \emph{M}-\emph{H} curves in Fig. 1(d).

When the thickness of inserted Cu varies from 3 nm to 10 nm, three changes emerge as follows: (1) $V_{\mbox{\tiny ISHE}}$  decreases gradually and even disappears due to increased spin relaxation in Cu \cite{Kimura2005} and decreased resistance of electrode B; (2) $V_{\mbox{\tiny PNE}}$  decreases because temperature gradient $\nabla T$ in NiFe also decreases; (3) $H_{\mbox{\tiny EB}}$ of NiFe increases with thicker Cu because the exchange coupling between NiFe and YIG weakens. On the other hand, once a 3 nm insulator MgO layer is inserted, $V_{\mbox{\tiny ISHE}}$  disappears while $V_{\mbox{\tiny PNE}}$ still exists under the same precision, as shown in Fig. 3(b), because thermal currents can still conduct even in insulators, but spin currents cannot. These results also confirm that the signal near 0 Oe is not from NiFe itself, such as ANE or PNE.

$T_{\mbox{\tiny B2,A}}$, $T_{\mbox{\tiny B3,A}}$ and $T_{\mbox{\tiny B2,C}}$, $T_{\mbox{\tiny B3,C}}$  represent the temperature of boundary 2, 3 of electrode B when heating electrode A and C respectively; $T_{\mbox{\tiny B2,A+C}}$, $T_{\mbox{\tiny B3,A+C}}$  represent the temperature of boundary 2, 3 of electrode B when heating A and C simultaneously. Due to the geometrical symmetry, $T_{\mbox{\tiny B2,A}}=T_{\mbox{\tiny B3,C}}$ , $T_{\mbox{\tiny B3,A}}=T_{\mbox{\tiny B2,C}}$, $T_{\mbox{\tiny B2,A+C}}=T_{\mbox{\tiny B3,A+C}}. V_{\mbox{\tiny ISHE}}$ and $V_{\mbox{\tiny PNE}}$ voltages satisfy the following equations:
 $V_{\mbox{\tiny ISHE,A}}=S_{\mbox{\tiny 2}}(T_{\mbox{\tiny B,A}}-T_{\mbox{\tiny 4,A}})=V_{\mbox{\tiny ISHE,C}}=S_{\mbox{\tiny 2}}(T_{\mbox{\tiny B,C}}-T_{\mbox{\tiny 1,C}})$, where $S_{\mbox{\tiny 2}}=\frac{1}{2}\theta_{\mbox{\tiny NiFe}}\eta_{\mbox{\tiny YIG-Cu-NiFe}}(L_{\mbox{\tiny NiFe}}/d_{\mbox{\tiny NiFe}})S_{\mbox{\tiny S}}$;
 $V_{\mbox{\tiny PNE,A}}=N(\textbf{\emph{M}})(T_{\mbox{\tiny B2,A}}-T_{\mbox{\tiny B3,A}})=-V_{\mbox{\tiny PNE,C}}=-N(\textbf{\emph{M}})(T_{\mbox{\tiny B2,C}}-T_{\mbox{\tiny B3,C}})$, where $N(\textbf{\emph{M}})$ is the simplified coefficient. When changing the heating electrode from C to A, $V_{\mbox{\tiny PNE}}$  is opposite in sign, while $V_{\mbox{\tiny ISHE}}$ is the same, as shown in Fig. 3(c).
When heating A and C at the same time:
$V_{\mbox{\tiny ISHE,A+C}}=S_{\mbox{\tiny 2}}(T_{\mbox{\tiny B,A+C}}-T_{\mbox{\tiny 4,A+C}})=S_{\mbox{\tiny 2}}(T_{\mbox{\tiny B,A+C}}-T_{\mbox{\tiny 1,A+C}})$,
$V_{\mbox{\tiny PNE,A+C}}=N(\textbf{\emph{M}})(T_{\mbox{\tiny B2,A+C}}-T_{\mbox{\tiny B3,A+C}})=0$. By eliminating  $\nabla T$ along \emph{y} axis in NiFe, $V_{\mbox{\tiny PNE}}$  in NiFe could nearly be cancelled, while $V_{\mbox{\tiny ISHE}}$ is enhanced because of the enhanced  $\nabla T$ in YIG. In this way, we succeed in directly detecting the pure $V_{\mbox{\tiny ISHE}}$  in NiFe without the influence of $V_{\mbox{\tiny PNE}}$ from itself [Fig. 3(c)]. Besides, $\nabla T$  along \emph{z} axis in NiFe will be also enhanced when simultaneously heating A and C. Even in this case, ANE voltages in NiFe are not observed, indicating that $\nabla T$ along \emph{z} axis in NiFe is negligibly small.

To further illustrate the ISHE in NiFe, we measured the $I_{\mbox{\tiny H}}$  dependence of $V_{\mbox{\tiny ISHE}}$ and $V_{\mbox{\tiny PNE}}$ , as shown in Fig. 4. The center field of the PNE curve corresponds to $H_{\mbox{\tiny EB}}$ of NiFe, and it decreases with increasing $I_{\mbox{\tiny H}}$ , as shown in Fig. 4(a), because $H_{\mbox{\tiny EB}}$ in FM/AFM usually decreases with the increasing temperature, even drops to zero at blocking temperature.

\begin{figure*}
\includegraphics[width=130mm]{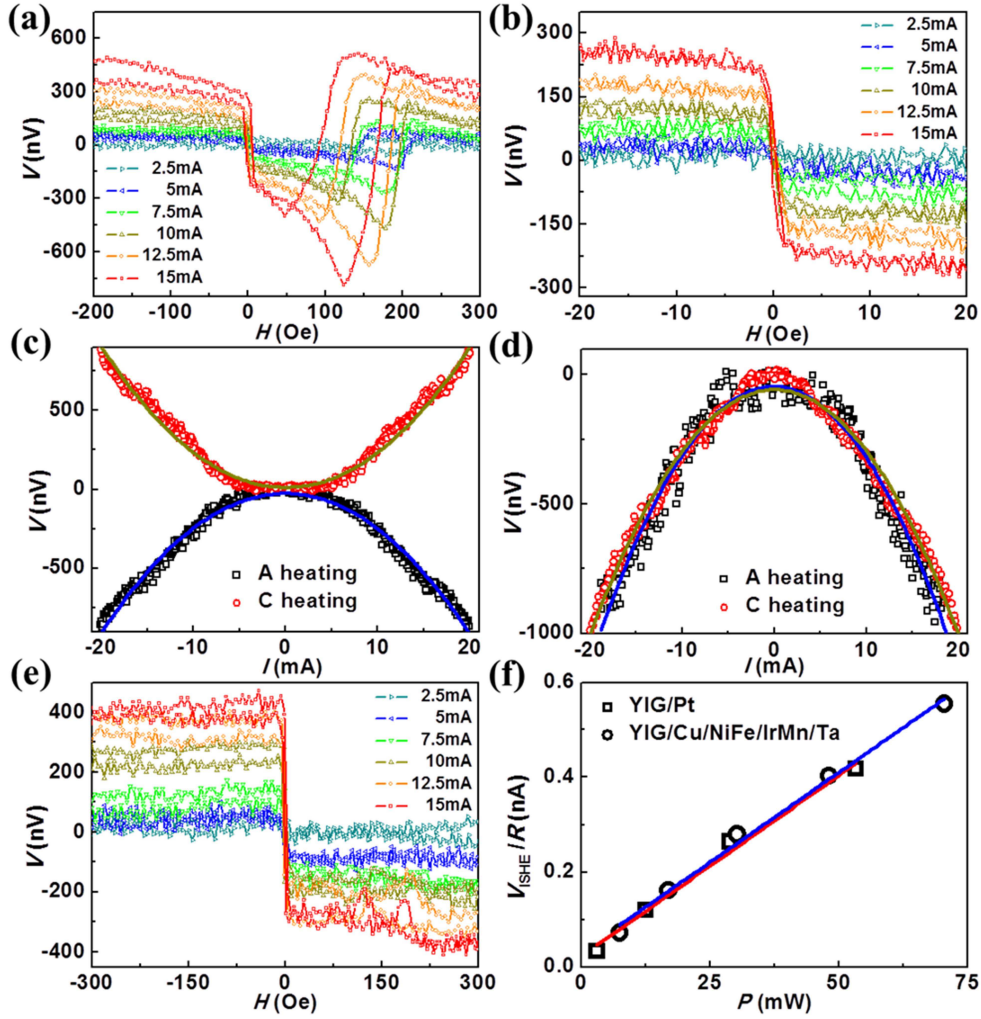}
\caption{\label{fig4}(a) - (e) were measured in YIG/Cu(5 nm)/NiFe(5 nm)/IrMn(12 nm)/Ta(5 nm) sample. We used the magnetic field range $\pm$1000 Oe and $\pm$50 Oe to measure the signal in (a) and (b) respectively. (a) shows the \emph{H} dependence of $V_{\mbox{\tiny ISHE}}$  and $V_{\mbox{\tiny PNE}}$  for different $I_{\mbox{\tiny H}}$  in electrode C, and (b) shows the pure \emph{H} dependent $V_{\mbox{\tiny ISHE}}$  due to the small field range. (c), (d) $I_{\mbox{\tiny H}}$ dependence and fitting curves of $V_{\mbox{\tiny PNE}}$ and $V_{\mbox{\tiny ISHE}}$ for heating electrode A or C respectively. (e) \emph{H} dependence of pure $V_{\mbox{\tiny ISHE}}$ for different $I_{\mbox{\tiny H}}$ in both electrodes A and C. (f) Heating power \emph{P} dependence of $V_{\mbox{\tiny ISHE}}/R$ in YIG/Pt(10 nm) and YIG/Cu(5 nm)/NiFe(5 nm)/IrMn(12 nm)/Ta(5 nm) samples.}
\end{figure*}

Fig. 4(c) and Fig. 4(d) show the $I_{\mbox{\tiny H}}$ dependence of $V_{\mbox{\tiny PNE}}$ [\emph{V}(+250 Oe)-\emph{V}(+20 Oe)] and $V_{\mbox{\tiny ISHE}}$ [\emph{V}(+10 Oe)-\emph{V}(-10 Oe)] respectively, they are both proportional to $I^2_{\mbox{\tiny H}}$ , confirming their thermal dependence. $V_{\mbox{\tiny PNE}}$  is opposite in sign when we changed the heating electrode from C to A, while $V_{\mbox{\tiny ISHE}}$  remains unchanged. This difference also confirms that these two signals should come from different origins: one from PNE in NiFe, and another from ISHE in NiFe induced by SSE in YIG. By simultaneously heating A and C, as shown in Fig. 4(e), enhanced pure $V_{\mbox{\tiny ISHE}}$  is observed, while $V_{\mbox{\tiny PNE}}$  from NiFe itself is totally eliminated.

To quantitatively analyze the spin Hall angle $\theta_{\mbox{\tiny SH}}$  of NiFe, we measured the $P$ dependence of  $V_{\mbox{\tiny ISHE}}$ in YIG/Pt(10 nm) and YIG/Cu(5 nm)/NiFe(5 nm)/IrMn(12 nm)/Ta(5 nm) samples, as shown in Fig. 4(f). ISHE induced charge currents: $V_{\mbox{\tiny ISHE}}/R=\beta\theta_{\mbox{\tiny SH}}P$ , where $R$ is the resistance of electrode B. We suppose the coefficient $\beta$ that expresses the efficiency from thermal currents to spin currents in electrode B is the same in these two samples. By linear fitting $V_{\mbox{\tiny ISHE}}/R-P$ curves, relative spin Hall angle $\theta_{\mbox{\tiny SH}}$(NiFe)/$\theta_{\mbox{\tiny SH}}$(Pt) $\approx 0.98$ . By using $\theta_{\mbox{\tiny SH}}$(Pt) $=0.1$ \cite{25wang2014scaling}, we obtain $\theta_{\mbox{\tiny SH}}$(NiFe) $=0.098$ , which is at the same order with $\theta_{\mbox{\tiny SH}}$(NiFe) $=0.02$  measured by spin pumping \cite{26wang2014spin}. These results show that NiFe almost has a comparable spin Hall angle with Pt. In fact, previous works have suggested strong SOC in 3d transition metals \cite{Du2014,Morota2011}and connected ISHE with AHE in the ferromagnetic metal (CoFeB) through Mott relation \cite{18wu2014unambiguous}. Strong SOC and ferromagnetic order in NiFe should contribute to the large $\theta_{\mbox{\tiny SH}}$. By using the exchange bias structure, investigating SHE and ISHE in ferromagnetic metals will become more feasible. As heavy metals with strong SOC, ferromagnetic metals become another promising candidate for detecting spin currents.

In conclusion, firstly a spin current in NiFe is generated by SSE in YIG, and then is detected by charge signals due to ISHE. The NiFe/IrMn exchange bias structure was used to separate ISHE and PNE in NiFe, and inserted Cu can decouple the exchange coupling and rule out the possible magnetic proximity effect between NiFe and YIG, allowing us to observe a pure ISHE signal. By simultaneously heating electrodes in both sides of electrode B, which can eliminate the in plane temperature gradient in NiFe, PNE from NiFe itself is eliminated, while only ISHE is remained. By fitting the $V_{\mbox{\tiny ISHE}}/R-P$ curves, we obtain a large spin Hall angle (0.098) in NiFe. This work is crucial to unambiguous confirmation of existence of ISHE in ferromagnetic metals and also to the applications of FM-based ISHE.

\begin{acknowledgments}
This work was supported by the State Key Project of Fundamental Research of Ministry of Science and Technology (MOST) [No. 2010CB934401], the MOST National Key Scientific Instrument and Equipment Development Projects [No. 2011YQ120053] and the National Natural Science Foundation [NSFC, Grant No. 11434014].
\end{acknowledgments}

%
\end{document}